\documentclass[12pt]{article}
\usepackage{esfconf}

\begin{document}


\title{Boundary Dynamics of Higher Dimensional Chern-Simons Gravity} 

\footnote{Invited talk given by G. Kunstatter at the International Conference "Quantization, Gauge Theory and Strings" dedicated to the memory of
Professor Efim Fradkin, Moscow, 2000.}


\authors{J. Gegenberg\adref{1} and  G. Kunstatter\adref{2}}


\addresses{\1ad Dept. of Math/Stats, U. of New Brunswick, Fredericton, NB, CANADA E3B 5A3,
\nextaddress
\2ad
Physics Department and Winnipeg Institute for Theoretical Physics, University of Winnipeg, Winnipeg, MB, CANADA R3B 2E9
}


\maketitle


\begin{abstract}
We review the relevance  to
the black hole entropy problem of boundary dynamics in Chern-Simons gravity. We then describe a recent derivation of the action  induced on the four dimensional
boundary in a five dimensional Chern-Simons gravity theory with gauge invariant, anti-deSitter
boundary conditions. 
\end{abstract}



\section{Introduction}

AdS spacetimes are under a great deal of scrutiny. One reason is their role in
 recent attempts to understand the
microscopic source of black hole entropy, as given by the Bekenstein-Hawking formula:
\begin{equation}
S_{bh} = {k A\over 4 G\hbar}
\end{equation}
A detailed understanding of the origins of this formula is still lacking, but
it is clear that it provides a remarkable mixture of geometry (the horizon 
area $A$), gravity (Newton's constant, $G$), quantum mechanics (Planck's constant $\hbar$)
and thermodynamics (Boltzmann's constant $k$). It is widely believed 
that the explanation for this formula need not necessarily be tied
to any particular theory of gravity, or its microscopic origin (such as string
theory). Instead a correct explanation might  apply equally well to any gravity theory that admits black hole solutions. 

One particularly elegant attempt at such a universal
 explanation was proposed in the mid-nineties (\cite{carlip96},\cite{balach95}). The idea was that black hole boundary conditions
caused gauge/diffeomorphism modes to become physical along the horizon, in analogy to the edge currents that appear in the Chern-Simons description of 
the Quantum Hall Effect. A concrete implimentation of this proposal was
given by Carlip~\cite{carlip96}, who considered Einstein gravity in 2+1 dimensions with
negative cosmological constant. This theory has a gauge theory formulation~\cite{gauge_grav}
as a Chern-Simons theory with group $SO(2,2)\sim SL(2,R))\times SL(2,R))$.
The corresponding action is:
\begin{equation}
S^{(3)}_{CS}= \int_{M^3} Tr(
A\wedge dA + {2\over 3} A\wedge A \wedge A)
\label{2+1 CS}
\end{equation}
The solutions to this theory describe spacetimes with constant negative 
curvature. One such solution is the BTZ black hole~\cite{btz} which can be
obtained from 2+1 dimensional
AdS spacetime by making suitable identifications. Carlip's argument went as follows: if one treats the black hole event horizon as a boundary then it is necessary to add  to the action a  surface term at the horizon in order to make the variational principle well defined. For example, in the case of the Euclidean 
BTZ black hole, the existence of a horizon requires,  in holomorphic coordinates ($z, \overline{z}$), that $A_z$ be fixed on the boundary (i.e. the horizon). The appropriate boundary term at the horizon is:
\begin{equation}
S_{bound}= \int_{\partial M^3} Tr(A_z A_{\overline{z}})
\end{equation}
The resulting total action $S_{total}=S^{(3)}_{CS}+S_{bound}$ is not gauge invariant. Under the transformation $A = g^{-1}\tilde{A} g + g^{-1}dg$, it changes by
\begin{equation}
\delta S_{total} = S_{WZW2}[g,\tilde{A}_z]
\end{equation}
where
\begin{equation}
 S_{WZW2}[g,\tilde{A}_z]={1\over 4\pi} \int_{\partial M}
    Tr(g^{-1}\partial_zg g^{-1}\partial_{\overline{z}}g
-2g^{-1}\partial_{\overline{z}} g  \tilde{A}_z)
 +{1\over 12\pi}\int_MTr(g^{-1}dg)^3
\end{equation}
This is the action for the gauged, chiral Wess-Zumino-Witten action in two dimensions (WZW${}_2)$~\cite{wzw2}. More correctly, an independent chiral WZW${}_2$ action emerges
for each copy of SL(2,R).
Carlip interpreted this as indicating that black hole boundary conditions caused
certain
  gauge modes (or diffeomorphism modes in the geometrical theory) on the boundary to become dynamical. By quantizing
the WZW${}_2$  boundary theory  and  counting states using methods in conformal
field theory Carlip~\cite{carlip96} was able to derive the correct Bekenstein-Hawking entropy for the BTZ  black hole.

An interesting, but to some extent 
puzzling,  variation of this scenario was
introduced  by A. Strominger~\cite{strom}.  He started from an old result of 
 Brown and Henneaux~\cite{brown} who showed that 2+1 AdS spacetime contained an asymptotic set of symmetries consisting of a pair of Virasora algebras. Note that the original Brown and Henneaux paper only derived the algebra. It was not until much later that Coussart, Henneaux and van Driel~\cite{couss} derived the boundary action and
corresponding dynamics that give rise to this algebra. The action was that of 
the WZW${}_2$ model, suitably restricted to give a Liouville theory on the
boundary. Strominger applied the Cardy formula to 
 the central charge  derived by Brown and Henneaux and was able  to count the asymptotic density of states for boundary conditions at infinity consistent with the presence of a black hole. Remarkably, Strominger's calculation yielded precisely the Bekenstein-Hawking entropy of the black hole.  It therefore seems that one could count black hole states either at infinity or the horizon.

The question we would like to address is: what happens in higher dimensions? The present discussion is based to  a large extent on work published in \cite{gk1}.

\section{Chern-Simons Gravity in 2n+1 Dimensions}

\
The Chern-Simons action Eq.(\ref{2+1 CS}) can readily be generalized to any odd 
dimension. In $2n+1$ dimensions it is of the form~\cite{zumino}:
\begin{eqnarray}
S_{CS}^{2n+1}[A] = \int {\cal L}^{2n+1}
\label{2n+1 action}
\end{eqnarray}
where ${\cal L}^{2n+1}$ is a $2n+1$-form defined by the requirement that its
exterior derivative take the form:
\begin{eqnarray}
d{\cal L}^{2n+1}= <(F[A])^n>
\end{eqnarray}
where $F[A] = dA+ {1\over 2} [A,A]$ is the field strength and the product of 
forms here  denotes a wedge product: e.g.  $(F[A])^2:= F[A]\wedge F[A]$, etc.
The angle brackets
$<...>$ denote a symmetric,  invariant n-linear form in the Lie algebra of $G$. It can
be verified that the following Lagrangian density satisfies the above
criterion:
\begin{eqnarray}
{\cal L}^{2n+1}= (n+1)\sum^n_{i=o}{n!\over (n+i+1)i!(n-i)!} <A^{2i+1} dA^{n-1}>
\label{2n+1 density}
\end{eqnarray}

In contrast to 2+1 dimensions, the higher dimensional 
 Chern-Simons theory does
have local, physical degrees of freedom~\cite{bgh}.  Moreover, the phase space is
stratified into ``layers'' of different dimension. In the generic sector, which
has maximal phase space dimension, there are $N-2$ dynamical modes for a gauge
group $G$ of rank $N$. 

Within the general class of theories described above, there is a subset that
has special importance for the present discussion. Consider the case in which
one has a gauge group $\hat{G}$  that, rather than being semi-simple, is a direct product
$\hat{G}=G\times U(1)$.
Banados {\it et al}~\cite{bgh} showed that Chern-Simons theory in $2n+1$ dimensions 
possesses an algebra of surface charges that is isomorphic to the algebra
for the $WZW_{2n}$ model. This is a $2n$ dimensional generalization of the
$WZW_2$ model. The action for the $2n$ dimensional model is\cite{iku}:
\begin{equation}
S_{WZW_{(2n)}}={i\over 4\pi}\int_M\omega^{n-1} \wedge \hbox{Tr}
(h^{-1} \partial h\wedge h^{-1}\overline{\partial} h)
+{i\over 12\pi} \int_{\partial M}\omega^{n-1}\wedge \hbox{Tr} (h^{-1} dh)^3
\label{wzw_2n}
\end{equation}
where $h$ is a field that takes its values in the group $G$ and 
$\omega$ is a  Kahler form that, in holomorphic coordinates
($z^\alpha,z^{\overline{\beta}}$), takes the form:
\begin{equation}
\omega = (i f_\pi^2/2) \omega_{\alpha\overline{\beta}}
   dz^\alpha dz^{\overline{\beta}}
\end{equation}
  $\partial,\overline{\partial}$ denote partial derivatives with respect to
the corresponding holomorphic coordinates $z^\alpha,z^{\overline{\beta}}$. The field equations that extremize the action Eq.(\ref{wzw_2n}) 
are:
\begin{equation}
\overline{\partial} (\omega^{n-1} \wedge h^{-1} \partial h) = 0
\end{equation}
or equivalently
\begin{equation}
{\partial} (\omega^{n-1} \wedge h^{-1} \overline{\partial} h) = 0
\end{equation}
  In the following section we will restrict our attention to $n=2$.
We will show that for 4+1 dimensional Chern
Simons gravity, the boundary action, dynamics and symmetry algebra 
are those of the WZW${}_4$ model. This latter model has been studied
by a variety of authors~\cite{wzw4}. Its field equations are equivalent to those of self-dual Yang-Mills theory in a particular gauge and the model is exactly solvable.
Moreover, it was shown by Ketov~\cite{ketov}, that the model is finite at one loop and it
is speculated that it may be finite at all orders. The symmetry algebra for
WZW${}_4$ is the so-called ``two-toroidal Lie algebra:
\begin{equation}
\{Q^a(x), Q^b(y)\} = {1\over 2} f^a_{cd} g^{bc} Q^d \delta(x-y)
  +{1\over 2} \epsilon^{ijk} \omega_{ij}\partial_k g^{ab} \delta(x-y)
\end{equation} 
It is a generalization of
the Kac-Moody algebra and has previously been studied by Mickelsson in the 
context of topologically massive Yang-Mills theory~\cite{mick1}.

\section{Boundary Dynamics 4+1 AdS Chern-Simons Gravity}
We start with the Chern-Simons action in 4+1 dimensions with gauge group
$\hat{G} = \hbox{SO}(4,2)\times U(1)$. From Eq.(\ref{2n+1 action}) and Eq.({\ref{2n+1 density}) we have:

\begin{equation}
S^5_{CS}=\int<\hat{A}\wedge d\hat{A}\wedge \hat{A}> +{2\over5} <\hat{A}^5> + {3\over2}<\hat{A}^3 d\hat{A}>
\label{4+1 action}
\end{equation}
where $\hat{A}$ is a one form that takes its values in the Lie algebra of $
\hat{G}$. In a suitable basis it can be decomposed:
$\hat{A} = A^iJ_i + a J_o$, where $J_a$, $i=1..15$ are the generators of $\hbox{SO}(4,2)$ and $J_0$ is the generator corresponding to $U(1)$. In this basis we can
define a symmetric trilinear form suitable for constructing the C-S action
Eq.[\ref{4+1 action}]:
\begin{eqnarray}
\left<J_i  J_j J_k\right> &=& \hbox{Tr}(J_i J_j J_k)\nonumber\\
\left<J_i  J_0 J_j\right> &=& \left<J_i  J_j J_0\right>= \hbox{Tr}(J_i J_j)
   \nonumber\\
\left<J_0  J_0 J_i\right> &=& \left<J_0  J_0 J_0\right>=0
\label{trilinear form}
\end{eqnarray}
The action, expressed in terms of the fields $A$ and $a$ is:
\begin{equation}
S^5_{CS}[\hat{A}]= S^5_{CS}[A] + 3\int_{M^5}Tr(AdA + {2\over3}A^3) \omega,
\label{action 2}
\end{equation}
The field equations are:
\begin{eqnarray}
\hbox{Tr}((F\wedge F+2F\wedge\omega)J_a)&=&0\\
\hbox{Tr}(F\wedge F)&=& 0
\end{eqnarray}
As mentioned above the solution space for this theory is much richer than
the 2+1 dimensional case. There are local dynamical degrees of freedom and  the phase space splits into ``strata'' of different dimensions. We will focus
on the ``generic'' sector of the theory, which has maximal number of degrees
of freedom. This sector contains the physically relevant class of  solutions:
\begin{eqnarray}
F[A]&=&0\nonumber\\
\omega[a]&=& \hbox{invertible but otherwise arbitrary}
\end{eqnarray}
The reason that these solutions are interesting is that they correspond
 to   AdS spacetime in the gravity theory.
In order to make this connection explicit, consider the 5-metric in {\it bein} form:
\begin{equation}
ds^2_5= \eta_{ab}e^a_\mu e^b_\nu dx^\mu dx^\nu
\end{equation}
and define SO(4,2) Lie-algebra valued connection one form by:
\begin{equation}
A= \left[\begin{array}{cc} w^a{}_b & {e^a}\\
          -{e_b}& 0 \end{array}\right]
\label{5d potential}
\end{equation}
where $\omega^a{}_b$ is the spin connection for the frame fields. It can then
be verified that the condition $F[A]=0$ in terms of the geometrical
fields implies that the torsion vanishes and that the metric is locally AdS. It has 
been shown~\cite{banados2} that locally AdS black holes (analoguous
to the BTZ black hole) exist in 4+1 dimensions, so the Chern-Simons gravity
theory under consideration does admit at least such black hole solutions.
In the following we are interested only in the asymptotics, and we will
therefore only impose the condition $F[A]=0$ at infinity. Whether or not
the theory admits black hole solutions that are AdS  only asymptotically is
still an open question currently under consideration.

The plan for deriving the boundary dynamics is roughly the following. We start with the action (\ref{action 2}) and require that on the boundary
$F[A]=0$ while $\omega[a]$ is a fixed, invertible, but otherwise arbitrary two form. Note that in contrast to
most previous work, the boundary conditions that we are imposing are {\it gauge invariant}, so that the gauge potentials are fixed on the boundary only up
to arbitrary gauge transformations. The next step is to figure out what boundary term must be added to the action in order
to make the variational principle well defined. The result is that  a suitable
boundary term does exist if  the gauge modes on the boundary
obey the WZW${}_4$ field equations. Thus, as in the 2+1 dimensional theory, the gauge modes on the boundary obey dynamical equations, and are therefore
physical. The corresponding boundary action is precisely that of
WZW${}_4$. We will now summarize
the calculations that  lead to this result. Details can be found in \cite{gk1}
and \cite{gk2}.

Consider the total action, including boundary term:
\begin{equation}
S_{tot} = S^5_{CS}[\hat{A}]+3\int_{\partial M}
   \hbox{Tr} (A_+ \wedge A_-)
\end{equation}
where 
$A_+=A_\alpha dz^\alpha$ and $A_-= A_{\overline{\alpha}}dz^{\overline{\alpha}}$
and ($z^\alpha, z^{\overline{\alpha}}$) are holomorphic coordinates on the
boundary $\partial M$ of the five dimensional manifold $M$. Direct calculation
reveals that the variation of the total action contains a piece in  addition to
the standard piece that vanishes when the bulk equations of motion are
satisfied:
\begin{equation}
\delta S_{tot} = \int_M (\hbox{Eq. of Motion})
   - 6 \int_{\partial M} (A_-\wedge \delta A_+)\wedge \omega
\end{equation}
 The second term in the above  seems to
imply that we have failed: the boundary term we have added
 does not totally elimate 
boundary variations, and according to standard lore, the variational 
principle for the bulk modes is still not well defined. However, we recall
that our boundary conditions imply that $A$ must be at least locally pure
gauge on $\partial M$, so that 
\begin{eqnarray}
A_+|_{bound} &=& (h^{-1} \partial_\alpha h) dz^\alpha\nonumber\\
A_-|_{bound} &=& (h^{-1} \partial_{\overline{\alpha}} h) dz^{\overline{\alpha}}
\end{eqnarray}
and  the boundary variation takes the form:
\begin{equation}
6\int_{\partial M}\hbox{Tr}(\partial(h^{-1} \overline{\partial} h \wedge \omega) \delta h)
\end{equation}
Thus the boundary term in the variation of the action vanishes if and only
if the gauge modes $h$ on the boundary satisfy  the WZW4 equations of motion:
\begin{equation}
\partial(h^{-1} \overline{\partial} h \wedge \omega)=0
\label{wzw4b}
\end{equation}

In order to verify that the proposed boundary term gives a consistent variational principle, we evaluate the 
full action for generic solutions to  the  bulk field equations, i.e. for
flat connection $A=h^{-1}dh$ and arbitrary $a$. The total action is
not gauge invariant, and  hence does not vanish for flat connections. Instead one gets:
\begin{equation}
S[h] = -\int_{M} Tr(h^{-1}dh)^3\wedge\omega+3 \int_{\partial M} Tr[(h^{-1}\partial
h)\wedge
     (h^{-1}{\bar \partial}h)]\wedge\omega,
\label{eq: wzw4 action}
\end{equation}
This is the WZW${}_4$ action, whose variation yields Eq.(\ref{wzw4b}) above. 

Note that physical, 
gauge invariant quantities in the bulk are not restricted by these extra
field equations, which only impose conditions on the boundary values of 
gauge transformations that can be performed
on the potentials.

The decomposition into gauge invariant bulk modes and the physical gauge
modes on the boundary can be made explicit using a parameterization first
used in \cite{bf}. The parametrization consists of
the usual Hodge decomposition for the abelian gauge potential:
\begin{equation}
a= a_h+d\lambda +\delta\beta
\end{equation}
In the above, $a_h$ is the harmonic part of the abelian potential $a$, $d$ is
an exterior derivative, $\delta$ is the corresponding co-derivative and
$\beta$ is a (gauge invariant) two form that determines the abelian field strength $\omega$, via the relationsip: $\omega= d\delta \beta$.
For the  non-Abelian gauge potentials, we use a generalized Hodge decomposition:
\begin{equation}
A= \tilde{A} + \delta_{\tilde{A}}\alpha^L
\label{parametrization}
\end{equation}
where $\alpha^L$ is a longitudinal two-form, $\tilde{A}$ is a flat connection that can itself be parametrized in terms
of global gauge transformations $h$:
\begin{equation}
\tilde{A}= h^{-1} \theta h + h^{-1} dh
\end{equation}
 with 
$\theta$ its topologically non-trivial part that cannot be globally gauge
transformed to zero. Moreover, $\delta_{\tilde{A}}$ is a covariant, co-exact
derivative with respect to the flat connection $\tilde{A}$. For details, and
motivation for this parametrization,  see \cite{bf}. The net effect of the above is to express an arbitrary field configuration in terms of a gauge
field $h$, a longitudinal, gauge covariant two form $\alpha^L$, and a topologically non-trivial flat connection $\theta$. In terms of this parametrization, 
the partition function looks like:
\begin{eqnarray}
Z&=& \int {{\cal D}A\over V_G}{{\cal D}a \over V_{U(1)}} e^{iS_{tot}} \nonumber\\
 &=& \int {\cal D}\theta J[\theta] Z_{WZW_4}[h_B,\theta]Z_{CS}[\beta,\alpha,\theta]
\label{partition}
\end{eqnarray}
Thus the partition function splits into a factor that is essentially the Chern-Simons partition function for the bulk modes with the given boundary conditions, and a factor that is the WZW${}_4$ partition function for the boundary modes. There is in addition an integral, with appropriate measure, over the topologically inequivalent flat connections, $\theta$ that the spacetime admits for the given boundary conditions. A detailed analysis of this
partition function will be presented in \cite{gk2}

\section{Conclusion}

We have shown that asymptotically AdS boundary conditions in 4+1 dimensional
Chern-Simons gravity lead to a WZW${}_4$ boundary action, in analogy with what happens in 2+1 dimensions. This talk started with the conjecture that such boundary terms will play an important role in our ultimate understanding of black hole entropy. The hope was to generalize the analyis of Carlip~\cite{carlip96} or Strominger~\cite{strom} to dimensions higher than three.  However, all we
 have so far is the boundary theory in 4+1 dimensions. We do not as yet know
how to count states, since the representation theory of the toroidal Lie algebras is not  well known, although some results have been reported by Billig~\cite{billig}. The other thing we would like to do is to extend the analysis to higher dimensions. A preliminary 
analysis seems to indicate that the boundary action is not precisely WZW${}_{2n}$ but this is currently under investigation as well.
\par
\vspace{5pt}
\noindent{\bf Acknowledgements.} The authors thank Y. Billig, M. Paranjape and
M. Visser for useful discussions and S. Carlip for pointing out to us the
alternative derivation of the boundary terms presented here.
G. Kunstatter is grateful to  M. Henneaux for useful comments
during the course of this conference and to the organizers for providing 
an excellent forum for discussion. This work is supported in part
by the Natural Sciences and Engineering Research Council of Canada.


\end{document}